\documentclass[review,12pt]{elsarticle}
\usepackage{lineno,hyperref}
\modulolinenumbers[5]

\journal{Journal of Templates}

\usepackage{amsmath}
\usepackage{graphicx}
\usepackage{natbib}

\title{Influence of external temperature gradient on acoustoelectric current in graphene}

\author[rvt]{K. A. Dompreh\corref{cor1}\fnref{fn1}}
\author[focal]{N. G. Mensah}
\author[rvt]{S. Y. Mensah}
\author[rvt]{R. Edziah}
\address[rvt]{Department of Physics, College of Agriculture and Natural Sciences, U.C.C, Ghana.}
\address[focal]{Department of Mathematics, College of Agriculture and Natural Sciences, U.C.C, Ghana}
\cortext[cor1]{Corresponding author\\ Email: kwadwo.dompreh@ucc.edu.gh\\
This work is licensed under the Creative Commons Attribution International\\ License (CC BY).
\url{http://creativecommons.org/licenses/by/4.0/}}
\ead[url]{kwadwo.dompreh@ucc.edu.gh}

\begin{document}

\begin{abstract}
Recent analyses of thermoelectric amplification of acoustic phonons in Free-Standing Graphene (FSG) $\Gamma_q^{grap}$ have 
prompted the theoretical study of the influence of external temperature gradient ($\nabla T$) 
on the acoustoelectric current $j_T^{(grap)}$ in FSG.  
Here, we calculated thermal field on open circuit ($j_T^{(grap)} = 0$) to be $(\nabla T)^g = 746.8Km^{-1}$. 
We then calculated acoustoelectric current ($j_T^{(grap)}$)to be $1.1mA\mu m^{-2}$ for $\nabla T = 750.0 Km^{-1}$, 
which  is  comparable  to that obtained in semiconductors ($1.0mA\mu m^{-2}$),
the thermal-voltage $(V_T)_0^g$ to be $6.6\mu V$ and 
the Seebeck coefficient $S$  as $8.8\mu V/K$.
Graphs of the  normalized  $j_T^{(grap)}/j_0$ versus $\omega_q$, $T$ and $\nabla T/T$ were sketched. For $j_T^{(grap)}$ on $T$
for varying $\omega_q$,  
Negative Difference Conductivity (NDC) ($\vert \frac{\partial j}{\partial T}\vert < 0$) was observed in 
the material. This indicates graphene is a suitable material for developing thermal amplifiers and logic gates.
\end{abstract}
\maketitle
\section*{Introduction}
The ability to acoustically generate d.c current in bulk and low dimemsional materials such as 
Superlattices (SL)~\cite{1,2,3,4}, Carbon Nanotubes (CNTs)~\cite{5,6,7,8} and Quantum 
wires (QW)~\cite{9,10} have recently become an active field of study. 
This phenomena is known as Acoustoelectric Effect (AE) and is caused by the attenuation of phonons  
leading to the appearance of a dc field. In Graphene, this effect has been verified 
theoretically~\cite{11,12,13} and experimentally~\cite{14,15,16,17}.
The high intrinsic carrier mobility (over $2\times10^5cm^2/Vs$) of a $2$-D graphene sheet,   coupled
with its amazingly high value for thermal conductivity at room temperatures 
($\approx 3000-5000 W/{mK}$), causes substantial acoustic effect when there is a minimal 
change in the external temperature gradient ($\nabla T$)~\cite{18}. This could lead to  activities such as 
AE~\cite{19}, amplification of acoustic phonons~\cite{20} or  Acoustomagnetoelectric effect 
(AME) in the sample~\cite{21,22}. The influence of non-linear thermal transport
in graphene has received little attention as against other non-linear effects such as 
electric and magnetic fields which are utilised in ideal atomic chains~\cite{24,25,26,27}, molecular
junctions~\cite{28} and quantum dots~\cite{29}. Daschewski et. al~\cite{33}, treated   the influence 
of energy density fluctuations (EDFs) on thermo-acoustic sound generation for near-field 
effects and sound-field attenuation for AirTech $200$, UltranGN-$55$ and thermo-acoustic transducer.
Hu et. al.~\cite{30} employed classical molecular
dynamics to study the non-linear transport in Graphene Nanoribbons (GNRs). The Negative
Differential Thermal Conductivity (NTDC) obtained by using the 
LAMMPS (Large-scale Atomis/ Molecular Massively Parallel Simulator) package and velocity
scaling software vanishes for lengths $ > 50nm$ long GNR. 
Such studies have  particular applications 
in thermal power sources such as thermophones, plasma firings 
and laser beams~\cite{30} but till date there is no  
theoretical study of the influence of  $\nabla T$  
on acoustoelectric effect in Graphene. 

In FSG, there are two types of phonons: ($1$) in-plane phonons with linear
and longitudinal acoustic branches (LA and TA); and (2) out-of-plane 
phonons known as flexural phonons (ZA and ZO)~\cite{32}.  
In this paper, we consider a stretched FSG  in which flexural phonons
are ignored and only in-plane phonons couples linearly to electrons. 
This study is done in the hypersound 
regime having $ql >> 1$ (where $q$ is the acoustic phonon wavenumber, $l$ is the electron mean-free path). 
Here, Negative Differential Conductivity (NDC) in FSG is reported. This is analogous to the 
electronic NDC~\cite{33,34} which is   a useful ingredient
for developing graphene based thermal systems such as signal manipulation 
devices, thermal logic gates and thermal amplifiers~\cite{31}.  

The paper is organised as follows: In the theory section, the equation underlying the 
acoustoelectric effect  in graphene is presented. In the  numerical analysis section,  the final equation is analysed  and 
presented in a  graphical form.   Lastly, the  discussion and conclusions are presented.    

\section*{Theory}
The acoustoelectric current ($j_T$) generated in a graphene sheet can be expressed as ~\cite{24,25}
\begin{eqnarray}
j_{{T}} = -\frac{e\tau A\vert {C_q}\vert^2}{(2\pi)^2 V_s}\int_0^\infty{kdk}\int_0^{\infty}{k^\prime dk^\prime}
\int_0^{2\pi}{d\phi}\int_0^{2\pi}{d\theta}\{[f(k)-f(k^\prime)]\times\nonumber \\
V_i\delta(k-k^\prime-\frac{1}{\hbar V_F}(\hbar\omega_q ))\} \label{Eq_1}
\end{eqnarray}
From ~\cite{26}, the matrix element $\vert C_q\vert$ in Eqn.($1$) is given as 
\begin{equation*}
\vert C_q\vert=
\begin{cases}{
\sqrt{\frac{\Lambda^2\hbar q^2}{2\rho \omega_q}} \hspace{92pt} \text{ acoustic phonons}}\\
{\sqrt{[(\frac{2\pi^2\rho\omega_0}{q^2})(k_{\infty}^{-1}-k_0^{-1})]} \quad \text{optical phonons}
}\end{cases}
\end{equation*}
where, ${\Lambda}$ is the constant of deformation potential,
$\rho$ is the density of the graphene sheet, $\tau$ is the relaxation constant, 
$V_s$ is the velocity of sound, $A$ is the area of the graphene sheet, 
$\omega_0$ is the frequency of an optical phonon, $k_{\infty}^{-1}$ and 
$k_{0}^{-1}$ are the low frequency and optical permeability of the crystal. 
The  linear energy dispersion  at the Fermi level with 
low-energy excitation is $\varepsilon(k) = \pm \hbar V_F \vert k \vert$ 
(the Fermi velocity $V_F \approx 10^8ms^{-1}$). From Eqn.($1$), the velocity $V_i$ is 
given as   $v(k)= {\partial\varepsilon(k)}/{\hbar\partial k}$ (where $V_i = v(k^\prime) - v(k)$ )
yields 
\begin{equation}
V_i =\frac{2\hbar\omega_q}{\hbar V_F}  \label{Eq_2}
\end{equation}
From Eqn.($1$), the linear approximation of the distribution function $f(k)$ is given as 
\begin{equation}
f(k) = f_0(\varepsilon(k)) +  f_1(\varepsilon(k))
\end{equation}
The unperturbed electron distribution function is given by the shifted Fermi-Dirac function,
\begin{equation}
f_0(k) = \{exp(\beta\varepsilon(k) - \beta\varepsilon_F) + 1\}^{-1}
\end{equation}
where $\beta = 1/k_{B}T$ ($k_B$ is the Boltzmann's constant and $T$ is the 
absolute temperature), and $\varepsilon_F$ is the Fermi energy. At low temperatures,
$\varepsilon_F = \xi$ ($\xi$ is the chemical potential) and the Fermi-Dirac equilibrium distribution function become 
\begin{equation}
f_0(\varepsilon(k)) = exp(-\beta(\varepsilon(k)-\xi)) 
\end{equation}
From Eqn. ($3$), $f_1(k)$ is derived from the Boltzmann transport 
equation as 
\begin{equation}
f_1(\varepsilon(k)) = \tau[(\varepsilon(k) - \xi)\frac{\nabla T}{T}]\frac{\partial f_0(p)}{\partial\varepsilon}v(k)
\end{equation}
Here   
$\tau$ is the relaxation time, and $\nabla T$ is the temperature gradient. With  $ k^\prime = k - \frac{1}{\hbar V_F}(\hbar\omega_q )$, 
and inserting Eqn.($2$), ($3$),($5$) and ($6$) into Eqn.($1$) and expressing further gives
\begin{eqnarray}
j_{{T}}=\frac{-e A\vert \Lambda\vert^2\hbar q\tau}{(2\pi)V_F\rho V_s}\int_0^\infty(k^2 -\frac{k\omega_q}{ V_F})
\{exp(-\beta(\hbar V_F k ))-\beta V_F q\tau(\hbar V_F k)\times\nonumber\\
 \frac{\nabla T}{ T}exp(-\beta\hbar V_F k)-exp(-\beta \hbar V_F (k - \frac{\omega_q}{V_F})) -
\beta \hbar {V_F} \tau(\hbar V_F(k -\frac{\omega_q}{V_F}))\times\nonumber\\
\frac{\nabla T}{ T}exp(-\beta \hbar V_F(k -\frac{\omega_q}{ V_F}))\}dk
\end{eqnarray}
Using standard integrals and after some cumbersome calculations,  Eqn($7$)
yields the  current ($j_T$)  as 
\begin{eqnarray}
j_{{T}} &=& j_0 \{(2 -\beta\hbar\omega_q)(1- exp(-\beta\hbar\omega_q))\nonumber\\
&-& \tau V_F [6(1+exp(\beta\hbar\omega_q)) -\beta\hbar\omega_q(2 + \beta\hbar\omega_qexp(\beta\hbar\omega_q))]\frac{\nabla T}{T}\}
\end{eqnarray}
where 
\begin{equation}
j_0 =  \frac{-2eA\tau\vert {\Lambda}\vert^2 q}{2\pi{\beta}^3{\hbar}^3{ V_F}^4\rho V_s}\label{Eq_6}
\end{equation}
From Eqn.($8$), for an open circuit ($j_T = 0$), the thermal field $(\nabla T)^{g}$ 
is  calculated as 
\begin{equation}
(\nabla T)^g = 
T\frac{\{(2 -\beta\hbar\omega_q)(1- exp(-\beta\hbar\omega_q))\}}
{\tau V_F [6(1+exp(\beta\hbar\omega_q)) -\beta\hbar\omega_q(2 + \beta\hbar\omega_qexp(\beta\hbar\omega_q))]}
\end{equation}
 the thermal field $(\nabla T)^g$ is found to depend on the temperature ($T$), the frequency 
($\omega_q$) and the relaxation time ($\tau$) as well as the acoustic wavenumber ($q$). The threshold 
temperature gradient $(\nabla T)^{g}$ relate 
the thermal voltage $V_T = k_{\beta} T/e$ as  
\begin{equation}
(\nabla V)_{T} =- S(\nabla T)^g
\end{equation}
where the Seebeck coefficient ($S$) is given as 
\begin{equation}
S = \frac{k_{\beta}\{\tau V_F [6(1+exp(\beta\hbar\omega_q)) -\beta\hbar\omega_q(2 + \beta\hbar\omega_qexp(\beta\hbar\omega_q))]\}}
{e(2 -\beta\hbar\omega_q)(1- exp(-\beta\hbar\omega_q))}
\end{equation}
\section* {Numerical Analysis}
To analyse, Eqn. ($8$), ($9$) and ($12$), we used the 
the following parameters: $\Lambda = 9 eV$, $V_s = 2.1\times 10^3ms^{-1}$, $\tau = 5\times 10^{-10}s$, 
$\omega_q = 10^{12} s^{-1}$ and ${q} = 10^4 m^{-1}$. At $T = 77 K$, the thermal field  generated  on 
open circuit $(\nabla T)^{g}$ is calculated to be $746.8 Km^{-1}$. 
To clarify the results obtained, the dependence of the normalized acoustoelectric current $j_{T} / j_0$ on  
$\omega_q$, $T$, $q$ and $\nabla T/T$ are  analysed graphically. 
In  Figure $1a$, the dependence of $j_T^{(grap)}/j_0$ on $\omega_q$ for 
varying $\nabla T$ are presented.  We observed that at $\nabla T = 850Km^{-1}$,
the graph rises to a maximum at $j_T^{(grap)}/j_0 = 2.8$ then decreased. 
By decreasing $\nabla T$ to $500Km^{-1}$, the graph decreases to a minimum at 
$j_T^{(grap)}/j_0 = -0.8$ and then increases.  
\begin{figure}[h!]
%\begin{centering}
\includegraphics[width = 7.5cm]{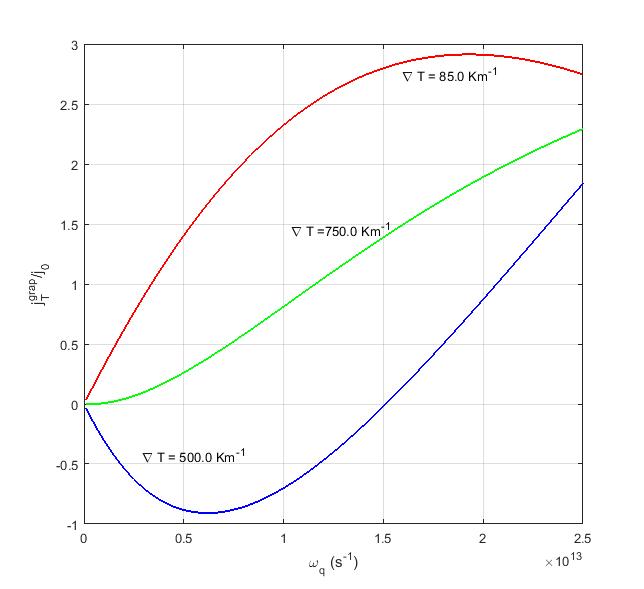}
\includegraphics[width = 8.0cm]{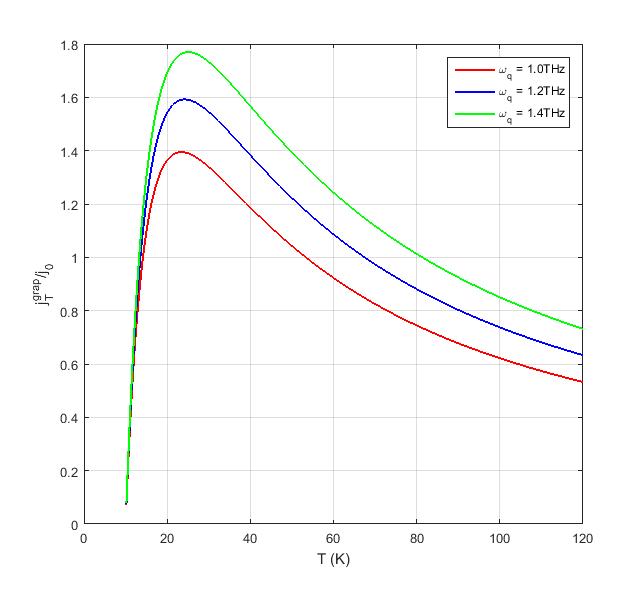}
\caption{(a) Dependence of $j_T^{(grap)}/j_0$ on $\omega_q$, (b) a graph of $j_T^{(grap)}/j_0$ on $T(K)$}
%\end{centering}
\end{figure}
Figure $1b$ shows the temperature dependence on the normalized acoustoelectric  current $j_T^{(grap)}/j_0$ for various 
$\omega_q$. We observed that for increasing temperatures, the graph raises to a peak value
and then decreases. The region of the decrease (negative slope ) indicates Negative Differential Conductivity (NDC)  
($\vert \frac{\partial j}{\partial T}\vert < 0$) in the materials.  The peak values increases with increases in $\omega_q$.
In Figure $2$, the behaviour  of $j_T^{(grap)}/j_0$ versus $\nabla T/T$ for varying $\omega_q$ and $q$ are 
presented. For Figure $2$, it was noted that the graphs initially attained  minimum points 
then increase for increasing $\nabla T/T$ to a maximum point then falls off. 
\begin{figure}[h!]
\center
\includegraphics[width = 8.5cm]{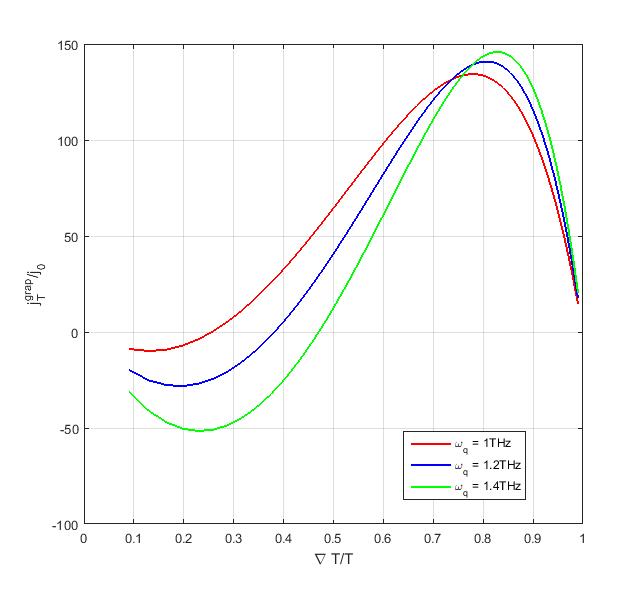}
\caption{ the dependence of $j_T^{(grap)}/j_0$ versus $\nabla T/T$ for varying $\omega_q$.}
\center
\end{figure}
It is observed that the ratio of the absolute value of the maximum peak $\vert j_T^{(grap)}/j_0\vert_{max}$ 
to the  minimum $\vert j_T^{(grap)}/j_0\vert_{min}$ peak is quite big. In the case where $\omega_q = 1.4THz$,
the ratio  $\frac{\vert j_T^{(grap)}/j_0\vert_{max}}{\vert j_T^{(grap)}/j_0\vert_{min}}  \approx 3$. A similar observation 
was made in superlattice for the case of electric field~\cite{4}. A $3D$ plot of the dependence  
of the normalized acoustoelectric current $j_T^{(grap)} /j_0$ on $\omega_q$ and $q$ are presented in Figure $3a$ and $b$. 
\begin{figure}[h!]
\includegraphics[width = 8cm]{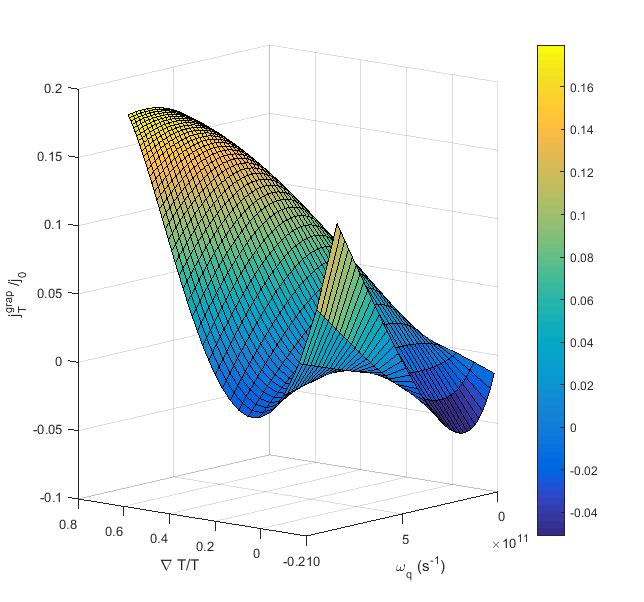}
\includegraphics[width = 8cm]{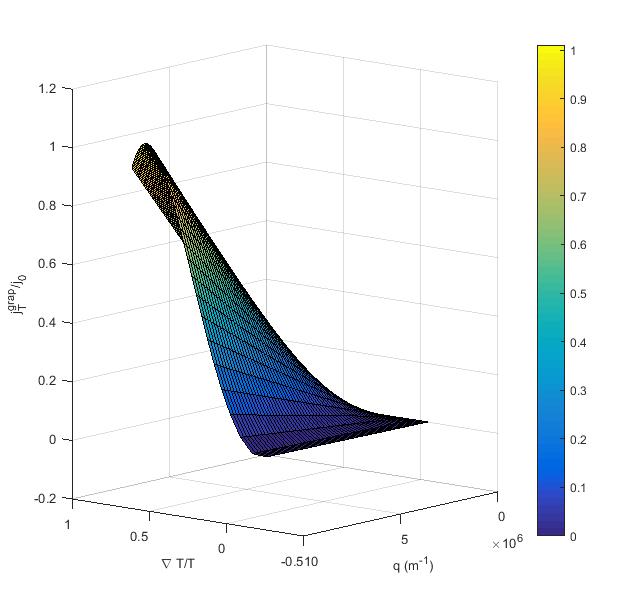}
\caption{(a) the dependence of $j_T^{(grap)}/j_0$ versus $\nabla T/T$ on $\omega_q$ (a)the dependence of $j_T/j_0$ versus $\nabla T/T$ on $q$ }
\end{figure}
The current density ($j_T^{(grap)}$)
generated per unit area in the sample at  $\omega_q = 0.1THz$ and 
$\nabla T = 750.0 Km^{-1}$
is calculated to be $j_T^{(grap)} = 1.1mA(\mu m)^{-2}$ as compared to  that calculated 
in semiconductors ($\approx 1.0mA(\mu m)^{-2}$).
Eqn.($12$) is the Seebeck coeffiecient $S$ which deals with the main 
thermoeletric properties of the FSG and how efficient it is. Fig. $4a$ 
shows the dependence of $S$ on $\omega_q$ for various $\nabla T/T$. The 
asymmetric  distribution 
is due to electrons moving at the Fermi level in the material with an 
energy related to the Fermi energy. The value of $S$ ranges from $152\mu V/K$  to 
$-22.7\mu V/K$ at $\nabla T/T = 0.16m^{-1}$, $215.5\mu V/K$ to $-322.7\mu V/K$
at $\nabla T/T = 0.22m^{-1}$,
and $278.8\mu V/K$ to $-417.6\mu V/K$ at $\nabla T/T = 0.29m^{-1}$ .
At $\omega_q > 2.16\times 10^{13}s^{-1}$, the graph switched from positive  
to negative values of $S$ indicating that at such frequencies, the n-type
FSG changes to p-type FSG. In Fig. $4b$,  the $S$ is plotted against $T$. Here,
the diffusion depends on temperature gradient present in the material which
creates the opposite field. From the graph,
the $S$ decreases with increasing T. At $\omega_q = 1.2\times 10^{12}s^{-1}$,  
and $T = 77K$, the  $S = 8.8\mu V/K$. By increasing the frequences also increases 
the value of the Seebeck coefficient. 
\begin{figure}[h!]
\includegraphics[width = 8.0cm]{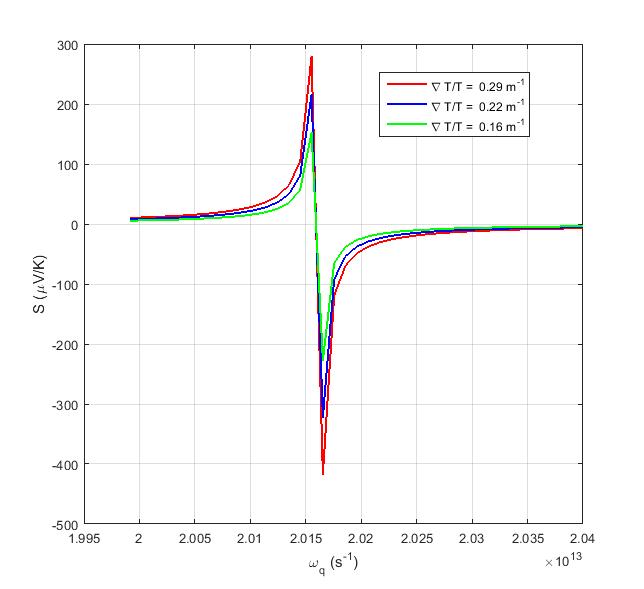}
\includegraphics[width = 7.5cm]{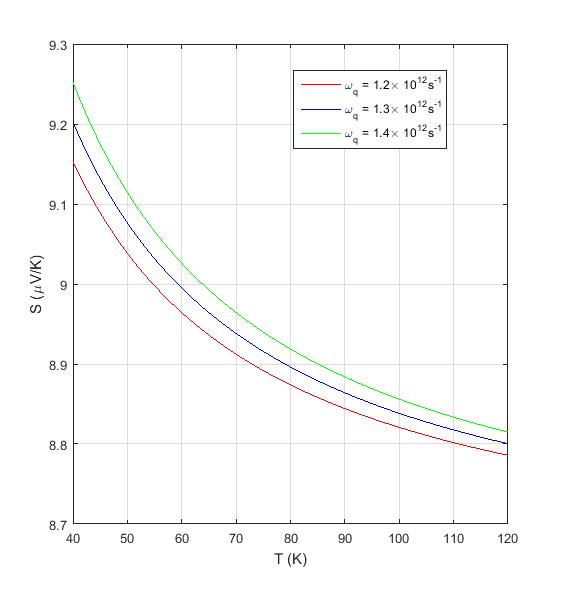}
\caption{(left) the graph of  $S$ versus $\omega_q$ for various $\nabla T/T$
(right)the dependence of $S$ versus $T$ for varying $\omega_q$}
\end{figure}

\section*{Conclusion}
The influence of external temperature gradient $\nabla T$ on  AE in FSG is studied. 
The thermal field $(\nabla T)^g$ is calculated to be $746.8Km^{-1}$. 
Negative differential conductivity ($\vert \frac{\partial j}{\partial T}\vert < 0$)  is observed to 
manifest in FSG. The  current density was calculated to be $j_T = 1.1mA\mu m^{-2}$ at $\omega_q = 0.1THz$ 
and the Seebeck coefficient evaluated to be $S = 8.8\mu V/ K$. FSG is therefore 
a suitable material for the development of thermal amplifiers and logic gates.

\section*{Reference}
% To change the title from References to Bibliography:
\renewcommand\refname{Bibliography}

\end{document}